\documentclass[]{aastex631}
\usepackage{graphicx}

\newcommand{\be}{\begin{equation}}
\newcommand{\ee}{\end{equation}}

\newcommand{\JADESeleven}{JADES-GS-z11-0~}

\newcommand{\JADESzthirteen}{JADES-GS-z13-0~}
\newcommand{\JADESfz}{JADES-GS-z14-0~}
\newcommand{\JADESfo}{JADES-GS-z14-1~}

\newcommand{\z}{z\simeq12}

\newcommand{\sigmav}{\langle\sigma v\rangle}

\begin{document}

\title{Neural Network identification of Dark Star Candidates. II. Spectroscopy}

\author{Adiba Amira Siddiqa}
\affiliation{Bryn Mawr College \\
101 N Merion Ave\\
Bryn Mawr, PA 19010, USA}

\author{Sayed Shafaat Mahmud}
\affiliation{Colgate University \\
 13 Oak Drive\\
Hamilton, NY 13346, USA}

\author{Cosmin Ilie}
\affiliation{Colgate University \\
 13 Oak Drive\\
Hamilton, NY 13346, USA}




\begin{abstract}
Some of the first stars in the Universe might be powered by Dark Matter (DM) annihilations, rather than nuclear fusion. Those objects, i.e. Dark stars (DS), offer a unique window into understanding DM via the observational study of the formation and evolution of the first stars and their Black Hole (BH) remnants. In \cite{NNSMDSPhot} (Paper~I) we introduced a feedforward neural network (FFNN) trained on synthetic DS photometry in order to detect and characterize dark star {\it photometric} candidates in the early universe based on data taken with the NIRCam instrument onboard the James Webb Space Telescope (JWST). In this work we develop a FFNN trained on synthetic DS spectra in order to identify {\it spectroscopic} dark star candidates in the data taken with JWST's NIRSpec instrument. In order to validate our FFNN model we apply it to real data for the four spectroscopic Supermassive Dark Star (SMDS) candidates recently identified in \cite{ilie2025spectroscopicsupermassivedarkstar} and reconfirm that indeed \JADESeleven, \JADESzthirteen, \JADESfz, and \JADESfo have spectra that are consistent with those of Supermassive Dark Stars. The main advantage of our FFNN model, in comparison to the Nedleaer-Mead Monte Carlo parameter estimator used in \cite{ilie2025spectroscopicsupermassivedarkstar}, is that the approach introduced here predicts parameters in milliseconds, over 10,000 times faster than the traditional method used in \cite{ilie2025spectroscopicsupermassivedarkstar}.
With this in mind, the FFNN model we developed and validated in this work will be adapted for Bayesian uncertainty analyses and automatic analyses of NIRSpec publicly available data for high redshift objects.
This study establishes a robust and efficient tool for probing Dark Stars and understanding their role in cosmic evolution.
\end{abstract}


\section{Introduction} \label{sec:intro}


Over the past decade, Neural Networks (NNs), a subset of Artificial Intelligence (AI) Machine Learning (ML) methods, have experienced an unprecedented surge in applications across diverse fields. These models leverage the power of pattern recognition to address complex real-world challenges~\citep[e.g.][]{samek2017explainable, choudhary2022recent, mienye2024comprehensive}. Deep learning, a specialized branch of machine learning, has revolutionized domains such as image classification \citep{wang2025s,oquab2023dinov2, ciregan2012multi, assran2023self}, text classification \citep{guleria2025nlp,sun2023text,minaee2021deep}, speech recognition \citep{nassif2019speech, sharrab2025advancements, ivanko2023review}, genome research \citep{yue2018deep, yue2023deep, gunduz2024optimized, barbadilla2025predicting}, and, more recently, astrophysics \citep{connor2018applying, walmsley2022galaxy, wei2020gravitational, auddy2022using, wong2021constraining, mahmud2025vader, zeraatgari2024exploring}. 

The application of machine learning (ML) in astrophysics has grown rapidly over the past decade, transforming how researchers analyze large and complex datasets. From classifying galaxies \citep{vavilova2021machine, cheng2020optimizing, fernandez2024detecting, zhong2024galaxy, ghaderi2025galaxy}, identifying exoplanets \citep{yu2019identifying, de2022identifying}, and detecting gravitational waves \citep{george2018deep, cuoco2020enhancing, koloniari2025new, de2024deep}, to mapping the cosmic web \citep{hong2021revealing, rodriguez2018fast, stuardi2025radio}, ML has proven to be a versatile and powerful tool for uncovering patterns in high-dimensional data. Among ML techniques, Feed Forward Neural Networks (FFNNs) have emerged as a cornerstone in astrophysics due to their ability to model highly non-linear relationships, enabling applications such as photometric redshift estimation \citep{razim2021improving, curran2021qso, zhou2025estimating}, lensing mass reconstruction \citep{gupta2021mass}, and supernova classification \citep{villar2020superraenn}. 

As the volume of data from the James Webb Space Telescope (JWST), ALMA, Euclid, or Vera Rubin Observatory (to name a few) continues to grow, the integration of ML in astrophysical workflows becomes increasingly essential. Neural networks, in particular, have been used to tackle complex problems involving noisy and incomplete datasets, making them invaluable for high-redshift studies \citep{stivaktakis2019convolutional, rastegarnia2022deep, zhang2025searching}.

The strength of NNs lies in their ability to train machines to recognize patterns and extract meaningful information through classification, detection, and regression tasks. In this study, we develop a Feedforward Neural Network (FFNN) model, trained on synthetic spectra of Supermassive Dark Stars (SMDSs). Our FFNN model is designed to infer crucial parameters such as mass and redshift of Dark Stars that can fit observed spectra taken with NIRSpec onboard JWST. One limitation of this work is that we do not include the possible role of nebular emission on Dark Stars, therefore assuming there is not sufficient H surrounding them to power a nebula. Our main goal here is to develop a tool for analysis of JWST NIRSpec data and fit it pure stellar SMDSs spectra, modelled with \texttt{TLUSTY}~\citep{hubeny2017TLUSTY}. In future work we plan to significantly extend the FFNN model used here, and include it into an automatic pipeline for JWST data analysis which will be able to handle both single isolated Dark Stars (with or without nebular emission) and Dark Stars embedded into early proto-galaxies. For more background on Dark Stars~\citep[e.g.][]{spolyar2008dark,freese2010supermassive,Ilie:2012}, observational status with JWST~\citep{Ilie:2023JADES,ilie2025spectroscopicsupermassivedarkstar}, and prospects of future detection with the Roman Space Telescope (RST)~\citep{Zhang:2022}, see the Introduction of the companion paper~\citep{NNSMDSPhot}. For a review on Dark Stars please consult \cite{Freese:2016dark}.

 In Paper~I~\citep{NNSMDSPhot} we introduced a FFNN model designed to handle JWST NIRCam data in order to identify Dark Star photometric candidates.  In turn, in this work we design of a FFNN that has the ability to handle JWST NIRSpec data quickly and reliably, in order to identify SMDSs spectroscopic candidates (see Sec.~\ref{sec:methods}). In Sec.~\ref{sec:results} we validate our FFNN models, and apply them to real data. In the process we confirm the results of \cite{ilie2025spectroscopicsupermassivedarkstar}, regarding the SMDSs spectral fits to each of the following objects: JADES-GS-z11, \JADESzthirteen, \JADESfz, and \JADESfo. The main advantage of the FFNN approach used here and in \cite{NNSMDSPhot}  is the extraordinary speed with which data is analyzed, thus allowing physical parameters of dark stars to be predicted far more efficiently than with the traditional methods used in previous works~\citep{Ilie:2023JADES,ilie2025spectroscopicsupermassivedarkstar}. Therefore, this work and the companion Paper~I~\citep{NNSMDSPhot} provide a foundation for integrating advanced machine learning models in an automatic pipeline we plan to design with the purpose identifying Dark Star candidates within large amount of archival JWST data (spectroscopic and photometric) from deep surveys dedicated to probing the high redshift universe. 
 
\section{Methods} \label{sec:methods}

\subsection{Neural Network Architecture} \label{subsec:NN_arch}

\begin{figure}[htbp]
    \centering
    \includegraphics[width=0.9\linewidth]{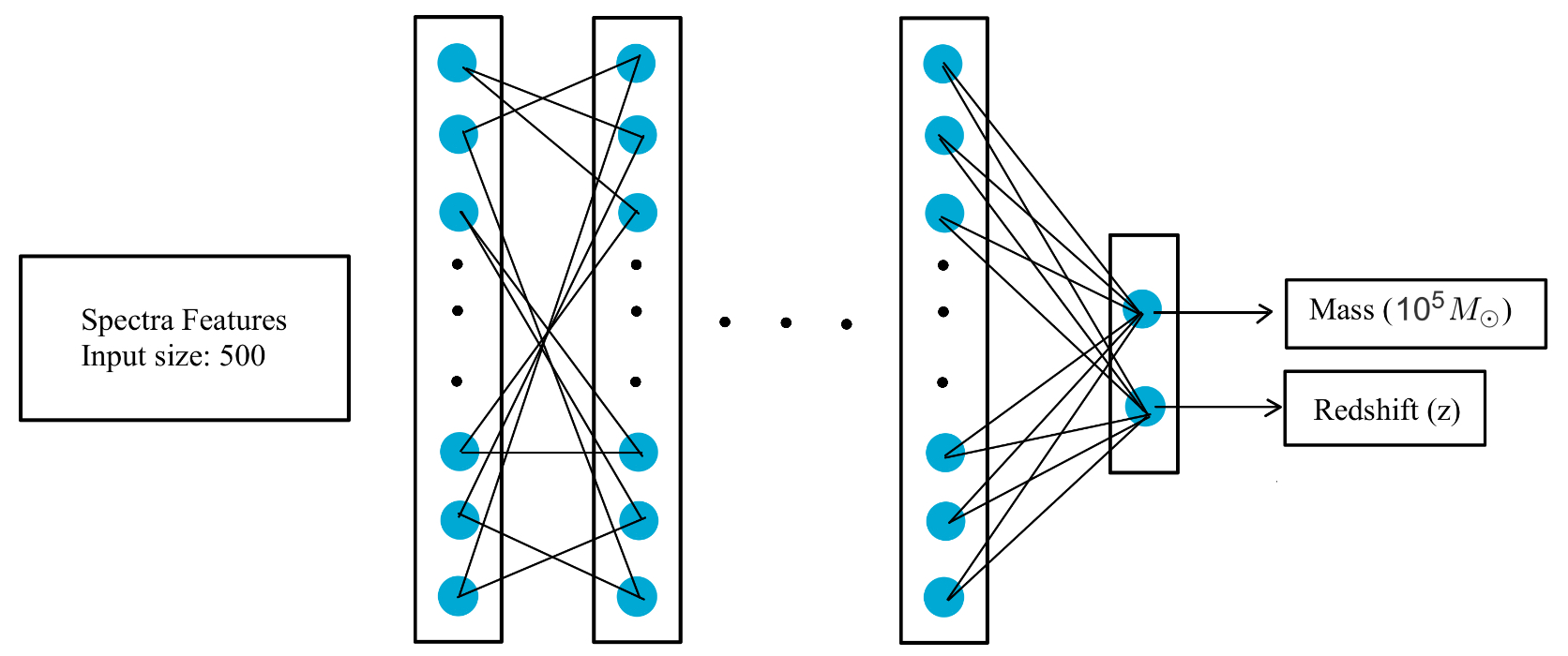}
    \caption{Architecture of the feed-forward neural network (FFNN) used to predict stellar mass and redshift from input spectra. The input layer consists of 500 flux values sampled uniformly across the 1–5.2\,\micron{} range. The network includes three fully connected hidden layers (256, 128, and 64 neurons) with ReLU activations, followed by a two-neuron output layer corresponding to mass (in units of \(\text{10}^5\,M_\odot\)) and redshift (\(z\)).}
    \label{fig:ffnn_spectra_architecture}
\end{figure}

As mentioned in the previous section, in this work we develop a feedforward neural network (FFNN) in order to identify Supermassive Dark Star candidates based on  observed JWST NIRSPec spectra for high-z objects, and estimate their Mass ($M$) and redshift ($z$). The input to the network consists of spectral flux values sampled at 500 evenly spaced wavelength points between 1 and 5.2 microns, as appropriate based on the sensitivity range of NIRSpec. The network architecture includes an input layer followed by three fully connected hidden layers containing 256, 128, and 64 neurons, respectively. Each hidden layer applies a rectified linear unit (ReLU) activation function \citep{agarap2018deep}, which introduces non-linearity and allows the network to learn more complex mappings between inputs and outputs. The final output layer has two neurons, one for mass and one for redshift, both treated as continuous variables.
The FFNN takes in the input spectrum and passes it through successive layers, each computing weighted combinations of the previous layer's outputs. These weights are adjusted during training to minimize discrepancies between the network's predictions and the known target values. The ReLU activation enhances the network’s ability to represent non-linear relationships \citep{goodfellow2016deep}, which are expected in the connection between spectral features and underlying physical parameters.
The spectral data used for training are generated from theoretical models of Supermassive Dark Stars formed via two different mechanisms: capture of dark matter particles and adiabatic contraction~\citep{freese2010supermassive}. We divide the dataset into training (72\%), validation (8\%), and test (20\%) subsets. To better simulate realistic observational conditions, we add 20\% Gaussian noise to the input flux values during training. This introduces variability that helps the network generalize to noisy observations.
We train the network using the Adam optimization algorithm \citep{kingma2014adam}, which adjusts learning rates adaptively to accelerate convergence and improve stability. For the capture-based models, we use a learning rate of $5\times10^{-4}$; for adiabatic contraction models, we use $1\times10^{-3}$. The loss function (Equation~\ref{eq:loss}) is the weighted average of the mean squared error (MSE), which quantifies the sum of average squared difference between predicted and true values of dark star mass and redshift. The goal of the neural network is to minimize this loss function. We assign a slightly higher weight to MSE for mass prediction ($w_M=2$) than that for redshift ($w_z=1$) because the effect of mass on the spectrum is more pronounced than the effect of redshift. We note that a lower MSE reflects better predictive accuracy.

\begin{equation}
\mathcal{L} = w_M \cdot \text{MSE}(M_{\text{pred}}, M_{\text{true}}) + w_z \cdot \text{MSE}(z_{\text{pred}}, z_{\text{true}})
\label{eq:loss}
\end{equation}

Training proceeds for 300 epochs with a batch size of 32. We use early stopping based on validation loss to prevent overfitting, halting training when performance no longer improves on unseen validation data.
After training, we assess model performance on the test set by computing the coefficient of determination ($R^2$ score) and the mean absolute error (MAE) for both mass and redshift. The $R^2$ score measures how well the model explains the variability in the target data, while MAE reports the average size of the errors in physical terms. We also produce scatter plots comparing predicted and true values to visually evaluate the accuracy and reliability of the model's output.
The FFNN is implemented in PyTorch \citep{paszke2019pytorch} and trained on a standard CPU. To ensure reproducibility, we save the network’s parameters whenever an improvement in validation loss is detected, retaining the best-performing model for evaluation. 

\subsection{Data Generation}
As was the case for Paper~I, we construct our dataset using simulated spectra produced with \texttt{TLUSTY}~\citep{hubeny2017brief}, a widely used code for modeling stellar atmospheres and generating theoretical spectra under non-local thermodynamic equilibrium (non-LTE) conditions. \texttt{TLUSTY} computes high-resolution spectral energy distributions (SEDs) for stars and compact objects, capturing both continuum and line features across a broad wavelength range. In this study, we use it to model the spectra of Supermassive Dark Stars (SMDSs) across a range of stellar masses.
We include two formation scenarios for SMDSs: capture and adiabatic contraction (AC), as described in detail in \citep{freese2010supermassive}. To generate a representative dataset for a given formation case, we uniformly sample 5000 mass values between $10^5$ and $10^7\ \mathrm{M_{\odot}}$, and 5000 redshift values $z\in[8-15]$, as this is the most likely interval for SMDSs to exist in. For each sampled mass–redshift pair, we compute the rest-frame SED and then redshift the spectrum, extracting the flux values over a set of 500 wavelength slices evenly spaced between 1 to 5.2 microns. 

For both formation scenarios, we apply a Lyman-$\alpha$ absorption cutoff to account for intergalactic medium attenuation at high redshifts, setting all flux to zero below the redshifted Ly$\alpha$ break. Additionally, Gaussian noise is added to each flux value to simulate observational uncertainties. The final dataset consists of arrays of flux values (sampled at fixed wavelength intervals), accompanied by the associated physical parameters: stellar mass and redshift.

\subsection{Model Performance}

\begin{figure*}[!htb]
    \gridline{\fig{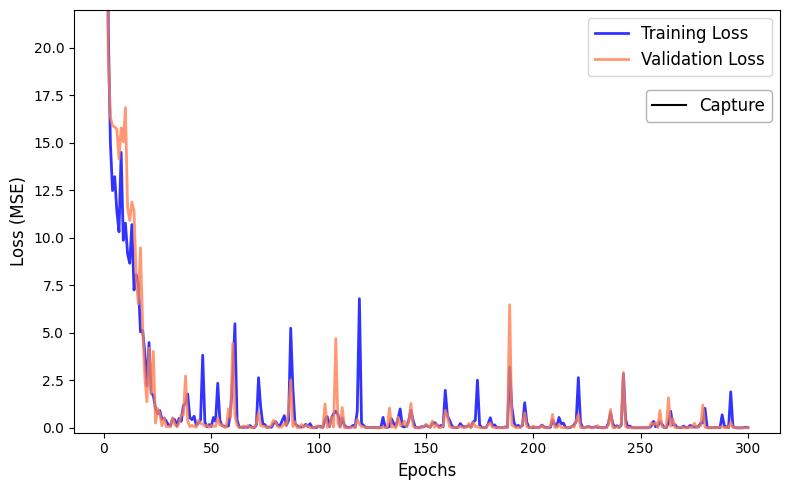}{0.48\textwidth}{(a)}
          \fig{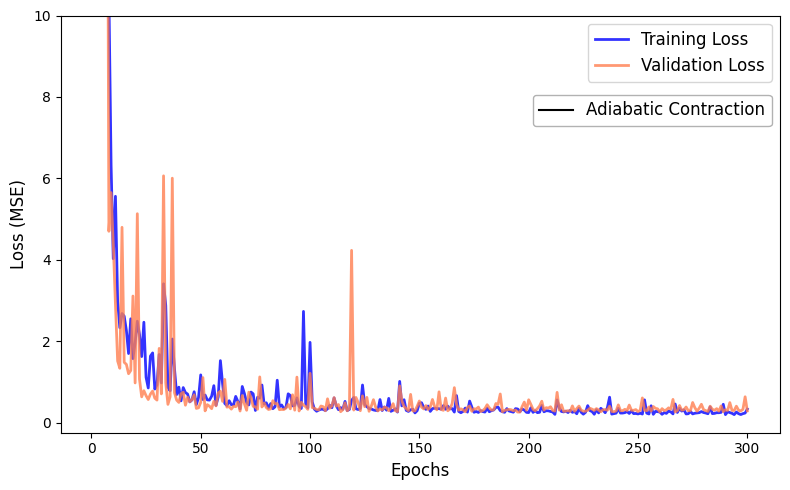}{0.48\textwidth}{(b)}}
    \caption{ Training and validation loss curves for the Feedforward Neural Network (FFNN) during the training process for the two scenarios of SMDSS formation: DM capture (panel a) and Adiabatic Contraction (panel b). The loss is measured in terms of Mean Squared Error (MSE) and is plotted against the number of epochs. The close alignment of the training and validation loss curves indicates the model generalizes well to unseen data without overfitting.}
    \label{fig:train-val-loss}
\end{figure*}

\begin{figure*}[!htb]
    \gridline{\fig{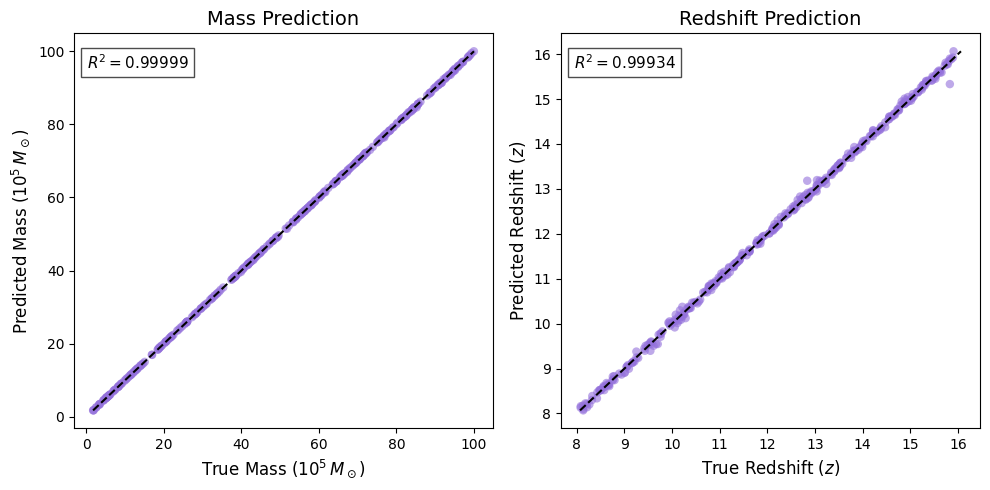}{0.45\textwidth}{(a)}
          \fig{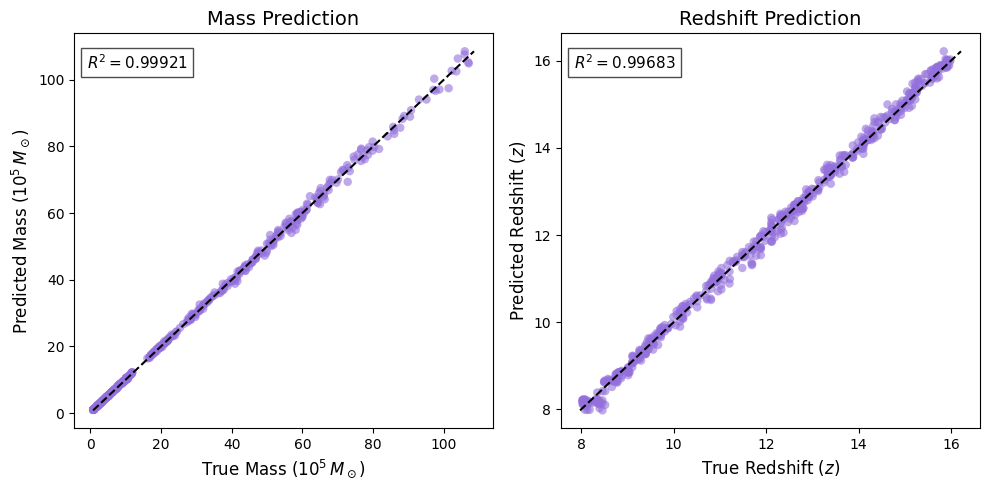}{0.45\textwidth}{(b)}}
    \caption{ Predicted versus true values for mass and redshift on test data for dark stars forming via DM Capture (panel a) and Adiabatic Contraction (panel b). The $R^2$ values indicate the model's performance, demonstrating excellent agreement between predictions and true values. The dashed red line represents the ideal $y=x$ relationship.}
    \label{fig:mass_redshift_predictionBoth}
\end{figure*}

We assess here (see Figs.~\ref{fig:train-val-loss}-\ref{fig:mass_redshift_predictionBoth}) the performance of the Feedforward Neural Networks (FFNN) using the test datasets for both the Capture and Adiabatic Contraction (AC) scenarios. The evaluation tests how well the model has been trained/validated (Fig.~\ref{fig:train-val-loss}) and the accuracy with which it recovers stellar mass and redshift from spectra (Fig.~\ref{fig:mass_redshift_predictionBoth}).

Both for SMDSs formed via DM capture and Adiabatic Contraction, the training and validation loss curves, shown in Figure~\ref{fig:train-val-loss} indicate stable convergence. The close alignment of the validation loss with the training loss suggests that the model generalizes well to unseen data and does not suffer from overfitting. Moreover, the predictions on the test sets for both SMDS formation scenarios (Figure~\ref{fig:mass_redshift_predictionBoth}) show strong agreement with the true values for both SMDS mass and redshift. The predicted points lie close to the one-to-one line (red dashed line), indicating high predictive accuracy. Our models achieve $R^2$ scores of $\gtrsim 0.99$ (both for mass and redshift), demonstrating near-perfect performance, even with Gaussian noise included in the training data.
Despite the added complexity introduced by injected noise, the model reliably predicts both parameters. 
Thus the FFNN designed here performs reliably for both formation channels, and in the next section we move on to applying it to real JWST NIRSpec data.

\section{Results}\label{sec:results}
\begin{figure*}[!ht]
\centering
\gridline{
  \fig{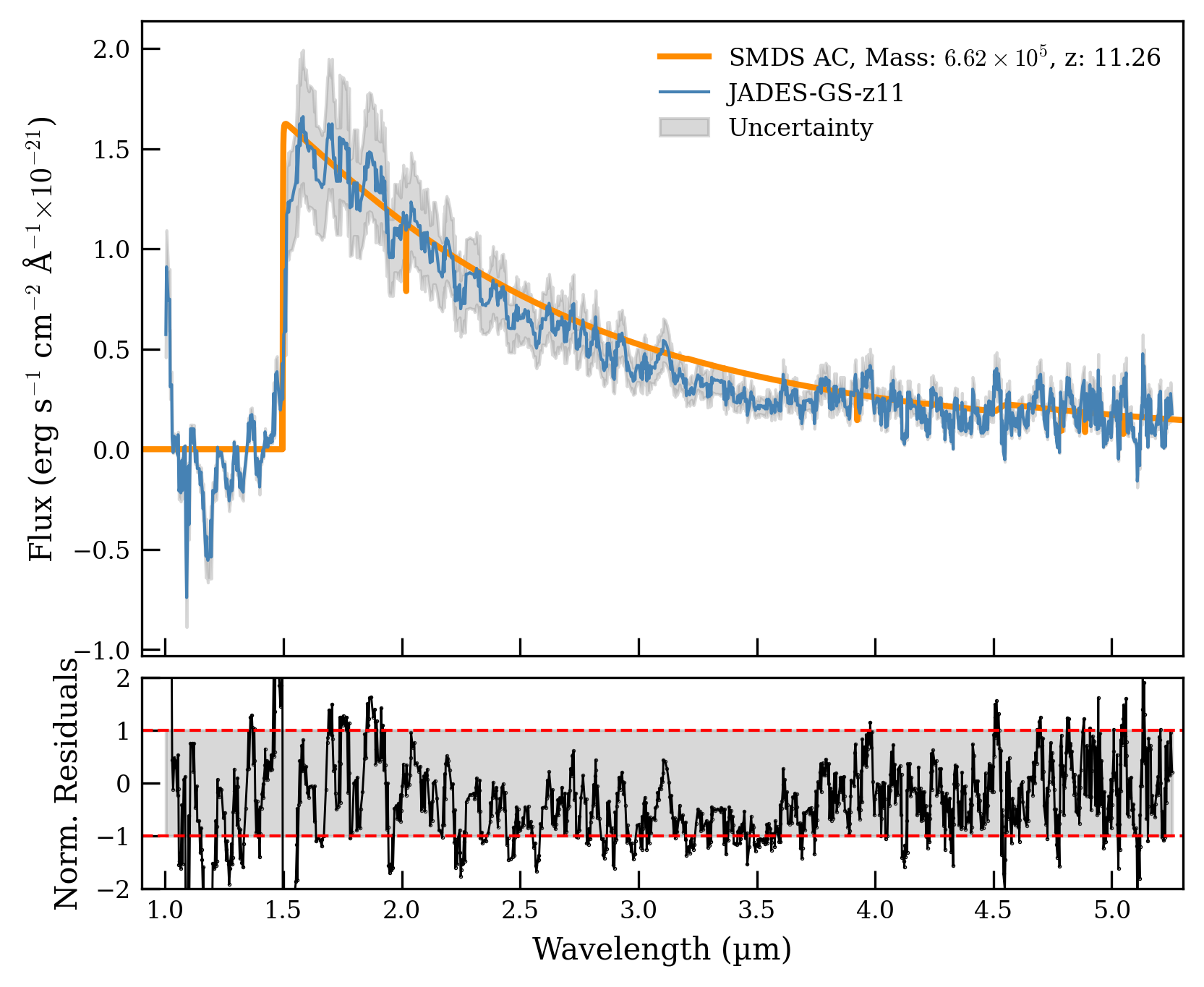}{0.45\textwidth}{\label{fig:ac_z11}(a) JADES-GS-z11}
   \fig{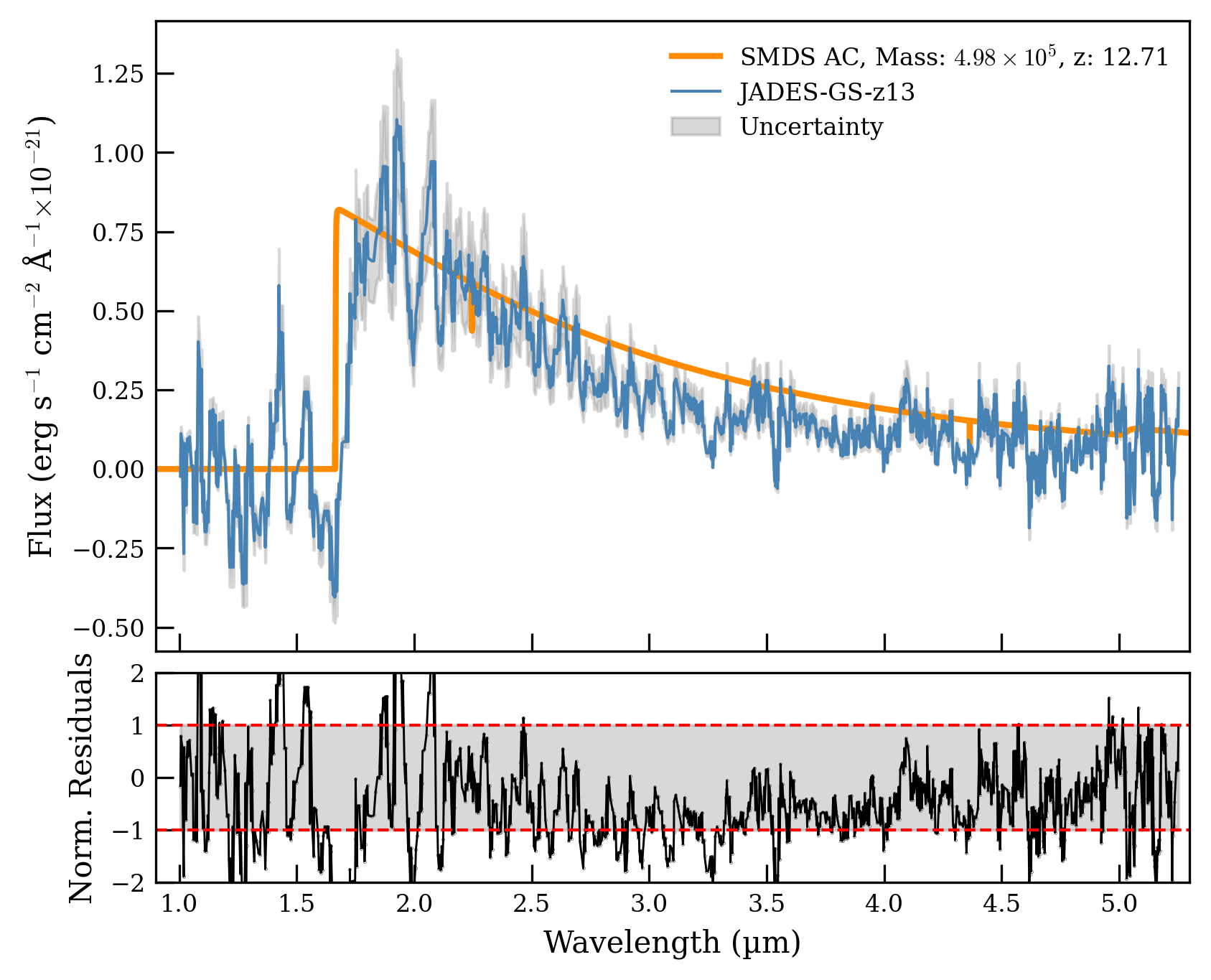}{0.45\textwidth}{\label{fig:ac_z13}(b) JADES-GS-z13}
}
\gridline{
\fig{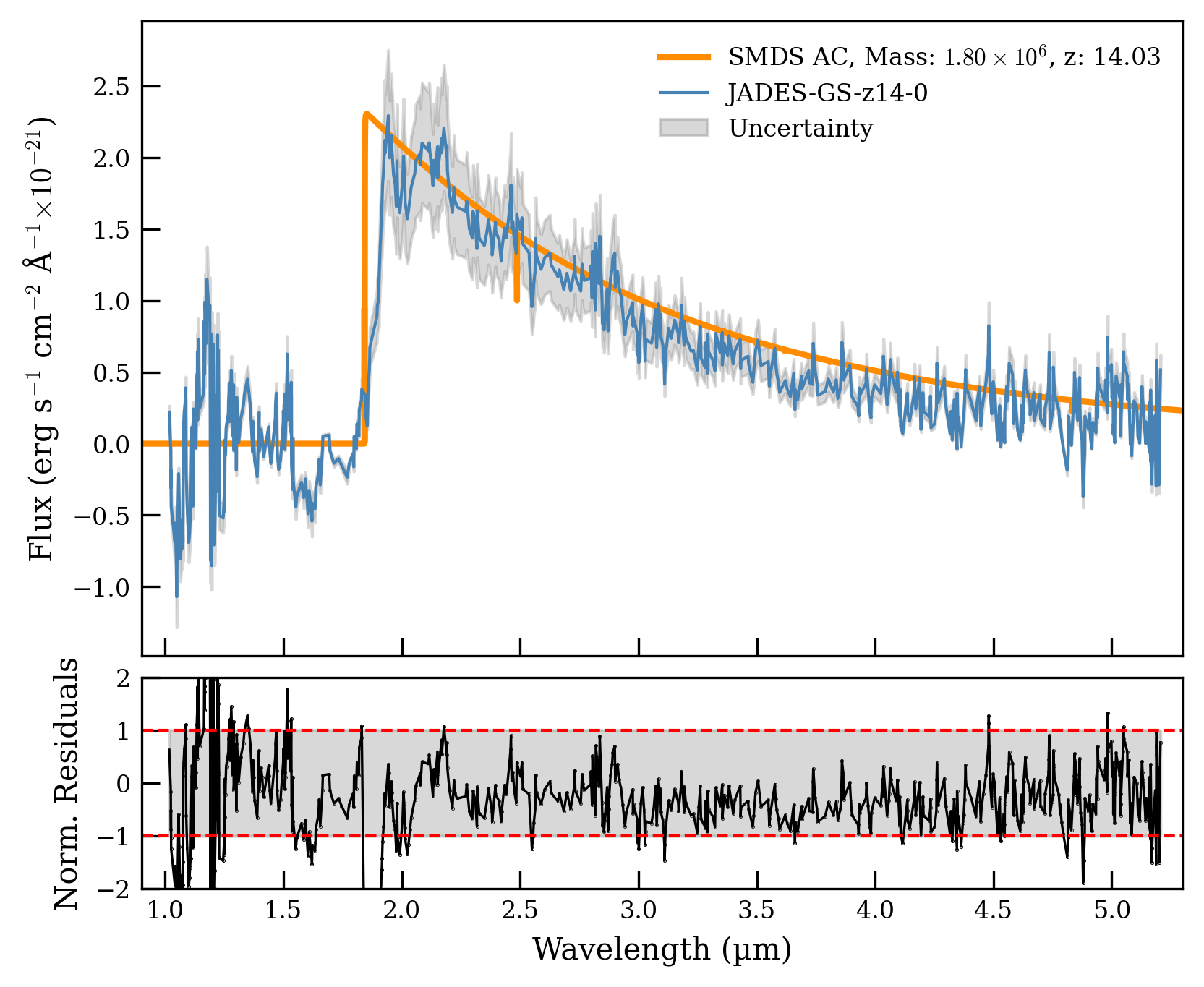}{0.45\textwidth}{\label{fig:ac_z14}(c) JADES-GS-z14-0}
\fig{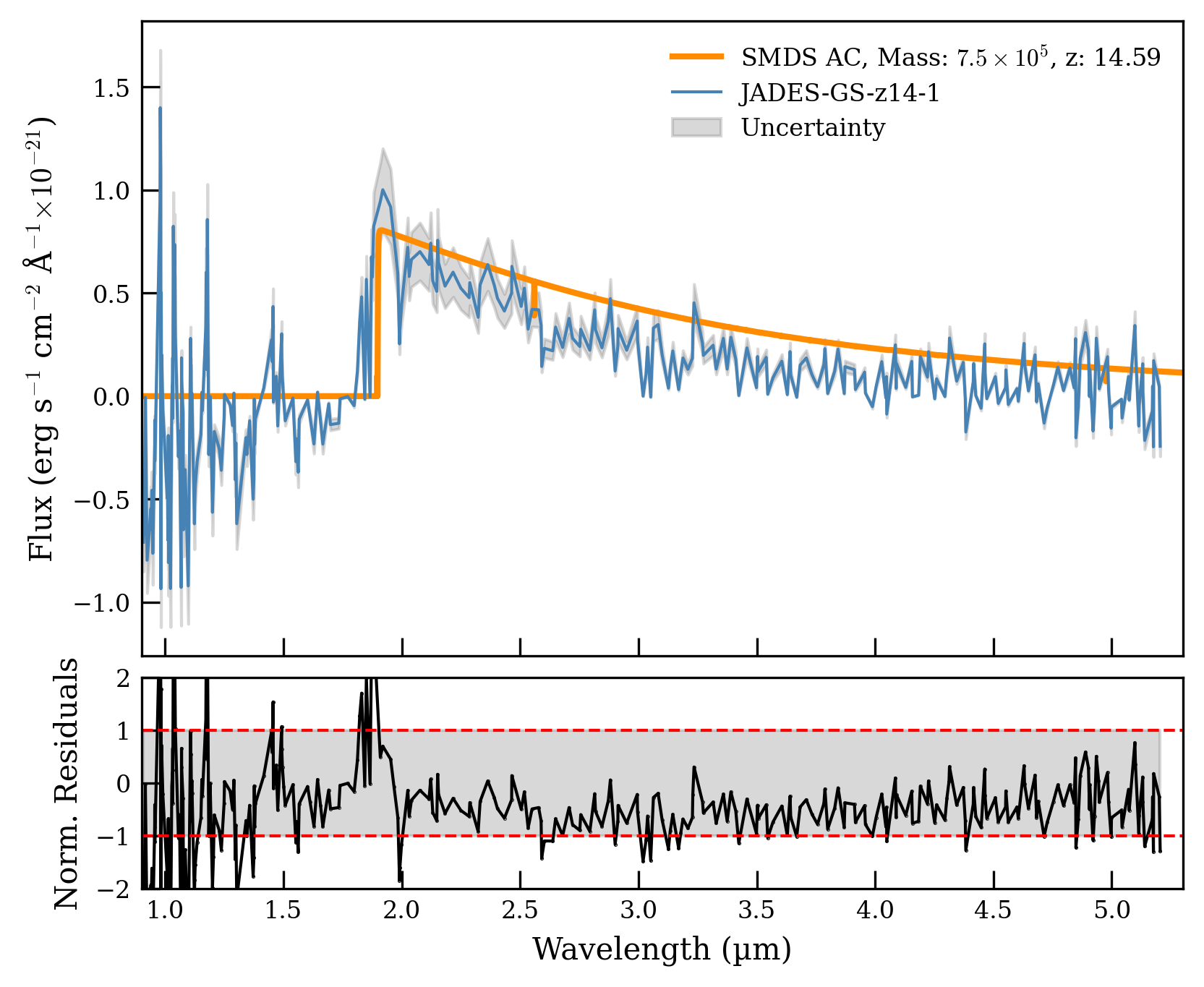}{0.45\textwidth}{\label{fig:ac_z14_1}(d) JADES-GS-z14-1}
}
\caption{
Comparison of FFNN model predictions with observed spectra for the four spectroscopic Supermassive Dark Stars (SMDS) candidates identified in \cite{ilie2025spectroscopicsupermassivedarkstar}.
Orange: FFNN model; blue: observed; gray: uncertainties. Residuals show (model–data)/$\sigma$ below each spectrum. 
}
\label{fig:combined_plots}
\end{figure*}

We select the data to analyze based our our previous results in \cite{ilie2025spectroscopicsupermassivedarkstar}, where the first SMDSs spectroscopic candidates were identified using a Monte Carlo parameter fitting routine based on the Needler-Mead optimization. In here we use our newly developed FFNN in order to check the findings of \cite{ilie2025spectroscopicsupermassivedarkstar}, and, in the process validate our new method. In Fig.~\ref{fig:combined_plots} and Table~\ref{tab:comparison} we present our main findings. 
Each panel in Fig.~\ref{fig:combined_plots} depicts a comparison of the the model-predicted spectra (orange line) with the observed data (blue line) and its associated uncertainty (gray shaded regions). Based on the normalized residuals (depicted with black in the sliver rectangular plots underneath each spectral fit) we conclude that our FFNN spectral fits demonstrate remarkable agreement between data and models, at the level of the continuum. In Table~\ref{tab:comparison} we present a comparison between the best fit parameters (SMDS Mass and redshift) obtained via our FFNN vs. those from \cite{ilie2025spectroscopicsupermassivedarkstar}. Generally the two models agree well, with some differences at the order of 10\% or less, which can be attributed to the fact that in this work we do not include any possible effects of nebular emission on the spectra of Dark Stars, whereas those are taken into account in \cite{ilie2025spectroscopicsupermassivedarkstar}, and their models had a third free parameter, the Hydrogen number density in a surrounding nebula. 

\begin{table*}[!ht]
    \centering
    \caption{Comparison of Neural Network Predictions and those made via Nedler-Mead MC approach in \cite{ilie2025spectroscopicsupermassivedarkstar}. The last column lists the computational time needed for the identification of each candidate with our FFNN.}
    \label{tab:comparison}
    \begin{tabular}{lccccr}
        \hline
        \textbf{Object Name} & \textbf{Mass ($10^5 M_\odot$)} (NN) & \textbf{Mass ($10^5 M_\odot$)} (MC) &\textbf{z} (NN) & \textbf{z} (MC) & \textbf{time} (NN) \\
        \hline
        JADES-GS-z11   & $6.62 $ & $6.35 $ &11.26 & 11.38 & 1.4 ms\\
        JADES-GS-z13   & $4.98 $ & $5.19 $ &12.71 & 13.20 & 1.78 ms\\
        JADES-GS-z14-0 & $18$ & $16.7$ &14.03 & 14.44 & 1.3 ms\\
        JADES-GS-z14-1 & $7.5 $ & $5.66 $ & 14.59 & 13.9 & 1 ms\\
        
        \hline
    \end{tabular}
\end{table*}



A major strength of our method is computational efficiency. Conventional spectral fitting for dark star detection, such as that employed in the earlier analysis by \cite{ilie2025spectroscopicsupermassivedarkstar}, involves generating theoretical stellar spectra with atmosphere codes like \texttt{TLUSTY} (and later \texttt{CLOUDY}\citep{CLOUDY:1998} for nebular corrections) and then performing spectroscopic comparisons with JWST/NIRSpec observations. Parameters such as stellar mass, effective temperature, redshift, and nebular density must be varied, and the best fit is found via iterative optimization (e.g., Nelder–Mead), repeated many times within Monte Carlo simulations to incorporate observational uncertainties. This procedure typically requires hours of computation for a single object. By contrast, our neural network predicts stellar mass and redshift in milliseconds—about $10^5$ times faster. This speed is especially critical for Bayesian posterior analyses, which demand thousands of repeated evaluations to propagate spectral uncertainties. While traditional approaches make such analyses prohibitively expensive, our model produces full posterior estimates almost instantly, enabling scalable application to large survey JWST/NIRSpec datasets, as detailed more in the next section.

\section{Limitations and Future Directions}\label{sec:Limitations}

Although the neural network introduced here predicts dark star model best fit stellar mass and redshif twith high accuracy, as demonstrated in the previous section, several limitations remain. First, the possible role of nebular emission from Dark Stars will be taken into account in future iterations of our FFNN algorithm, such that we can also couple it to the morphological analysis pipeline described in \cite{ilie2025spectroscopicsupermassivedarkstar}.~\footnote{\cite{ilie2025spectroscopicsupermassivedarkstar} model four of the dustless, metal poor, ultra-compact very bright high-z JWST galaxies (known as ``blue monsters''~\citep{ferrara2025blue}) as SMDSs powering a nebula, and as alluded to before, find compelling SMDSs fits both in terms of morphology and spectra to JWST data for \JADESeleven, \JADESzthirteen, \JADESfz, and \JADESfo.}  Moreover, the spectra used in this study are accompanied by associated errors, but our model does not fully incorporate this uncertainty into its predictions. To better capture how uncertainties in dark star spectra affect the inferred masses and redshifts, and to reflect the probabilistic character of neural networks, future work will investigate Bayesian Neural Networks (BNNs). A Bayesian framework would allow us to quantify uncertainties more rigorously by producing posterior distributions for the predicted parameters, rather than point estimates. Such improvements are crucial for maintaining model reliability, particularly at high redshift, where observational uncertainties become significant.

Another limitation is the rather small parameter space of dark stars considered in this study: Mass, redshift, and formation mechanism (Adiabatic Compression vs DM Capture). Expanding this feature space is critical for capturing the full diversity of potential dark star properties. For instance, embedding dark stars within a host galaxy and examining how the surrounding stellar and interstellar environments modify the spectrum could yield more realistic models. Moreover, all SMDSs models used so far in the literature to identify Dark Star candidates were based 100 GeV WIMPs, annihilating with the canonical cross section that leads to the observed relic abundance of DM: $\sigmav\sim 3\times 10^{-26}$cm$^3$s$^{-1}$. We plan to extend the range of WIMP masses explored. As such, once a statistically significant sample of spectroscopic SMDSs candidates has been identified, we would gain insights into the nature of the DM particle, via estimations of its most likely mass and annihilation cross section.  

All previous analyses conducted in order to identify Dark Star candidates were performed iteratively, one object at a time, and they were based on the rather slow Nedler-Mead optimization algorithm. As such, the amount of NIRCam/NIRSpec data analyzed in \cite{Ilie:2023JADES}/\cite{ilie2025spectroscopicsupermassivedarkstar} was very limited. Moreover, no MIRI data was yet used in those analyzes. So, in order to further test the Dark Star hypothesis, we will create a pipeline based on the extremely fast, yet robust FFNNs described this work (and Paper~I) that is able to automatically and efficiently handle the large amounts of high-redshift Universe archival JWST spectroscopic (and photometric) data available either in the MAST catalog~\citep{MASTPortal}, or from dedicated survey websites such as: CEERS~\citep{ceers2025}, JADES~\citep{jades_dr4_2025}, GLASS~\citep{glass_jwst_hlsp_2023}, UNCOVER~\citep{uncover_data_releases_2024}, and PRIMER~\citep{primer_survey_2023}. As of now, in those catalogs combined there are more than 5,000 objects with $z_{phot}\gtrsim 9$, and for about 50 this is confirmed with $z_{spec}$ estimated via the identification of a Lyman break with NIRSpec. The large amounts of relevant data is  motivating the creation of the pipeline we propose, and the work presented here and in Paper~I provides a cornerstone in this direction. 

Once a sufficiently large size of Dark Star candidates has been identified we will select the most promising ones for followup observations and target their smoking gun signatures, such as the He~II 1640{\AA} absorption feature already hinted at  ($S/N\sim 2$) by NIRSpec spectra of \JADESfz~\citep{ilie2025spectroscopicsupermassivedarkstar}. The confirmation of even a single Supermassive Dark Star, based on this, or potentially other, distinct spectroscopic signatures, would have enormous impact on astronomy, astrophysics, and particle physics. It would indirectly verify the existence of self-annihilating DM particles, while, at the same time, opening up a new field of Astronomy and Astrophysics: the observational study of Dark Matter powered stars.

\bibliography{sample631,RefsDM}{}
\bibliographystyle{aasjournal}



\end{document}